\documentclass[aps,prl,twocolumn,a4paper,10pt]{revtex4-1}

\usepackage[T1]{fontenc}		
\usepackage[english]{babel}		
\usepackage[latin1]{inputenc}	
\usepackage{times}

\usepackage{amsmath}			
\usepackage{amssymb}			
\usepackage{bbm}				
\usepackage{mathtools}
\usepackage{simpler-wick}
\usetikzlibrary{tikzmark}
\DeclarePairedDelimiterX\Dirbraket[3]{\langle}{\rangle}%
{#1\,\delimsize\vert\,\mathopen{}#2\,\delimsize\vert\,\mathopen{}#3}

\usepackage[colorlinks=true, a4paper=true, pdfstartview=FitV,linkcolor=blue, citecolor=blue, urlcolor=blue]{hyperref}

\usepackage{graphicx}			
\usepackage{graphics}			
\DeclareGraphicsExtensions{.pdf}
\usepackage{color}	

\newcommand{\bea}{\begin{eqnarray}}
\newcommand{\eea}{\end{eqnarray}}

\newcommand{\bs}{\boldsymbol}

\newcommand{\bk}{\mathbf{k}}
\newcommand{\bq}{\mathbf{q}}
\newcommand{\bsigma}{\boldsymbol{\sigma}}

\newcommand{\be}{\begin{equation}}
\newcommand{\ee}{\end{equation}}
\newcommand{\bpm}{\begin{pmatrix}}
\newcommand{\epm}{\end{pmatrix}}

\newcommand{\bA}{\mathbf{A}}
\newcommand{\bB}{\mathbf{B}}

\begin{document}

\title{Stability of line-node semimetals with strong Coulomb interactions and properties of the symmetry-broken state}

\author{Carlos~Naya}
\author{Tommaso~Bertolini}
\author{Johan~Carlstr\"om }
\affiliation{Department of Physics, Stockholm University, 106 91 Stockholm, Sweden}
\date{\today}

\begin{abstract}
We employ diagrammatic Monte Carlo simulations to establish criteria for the stability of line-node semimetals in the presence of Coulomb interactions. Our results indicate a phase transition to a chiral insulating state that occurs at a finite interaction threshold which we determine.  
We also compute the Landau levels for out-of-plane and in-plane magnetic fields in the symmetric and symmetry-broken phases. We find that the magnetic field couples to the chiral order parameter, implying that this degree of freedom can be manipulated in situ in experiments. 
Finally, we check the existence of edge states in the symmetry-broken phase. On the system's boundary, we note that the metallic "drum-head" states that exist in the symmetric phase are gapped out. However, the symmetry-broken phase permits topological defects in the macroscopic order parameter in the form of domain walls, which host metallic "interface states." These consist of line-like gap-closings that occur on the two-dimensional interfaces. 
\end{abstract}
\maketitle


Topological semimetals exhibit a band structure that is gapped everywhere except at a few nodal points, where the bands meet. The topological nature of these protects them against perturbations and also gives rise to unique phenomena like the chiral anomaly--where nodal points act as sources and sinks of a spontaneous current--and Fermi arcs, which are direct manifestations of the bulk topology \cite{Jia2016}.
Applications of topological semimetals are thus far mainly as building blocks of information technology \cite{Wang2020,Ma2019,Han2018}. Several of these materials exhibit a large magnetoresistance effect \cite{Kumar2017} that may be exploited in magnetic field sensors \cite{Wang2016}, and spintronic devices \cite{PhysRevLett.117.146403}. 
 
In the search for new and technologically useful semimetals, symmetry-protected topological phases have emerged as an important platform that dramatically expands the types of nodal features realized in a material. Typically, this involves a combination of point group symmetries and explicitly broken time-reversal or inversion symmetry. The result is a wide class of band touching points that carry more than unit topological charge \cite{Singh2018,PhysRevLett.108.266802} and may also involve multiple bands \cite{BeyondDirac, PhysRevLett.119.206402} or have to be classified as line-nodes \cite{Bzdusek2016,PhysRevB.96.155105,Bian2016,PhysRevB.93.121113,doi:10.7566/JPSJ.85.013708,doi:10.1063/1.4926545,2017Tanaka}.

While the explicit reliance on symmetry significantly widens the scope of topological and semimetallic materials, it also 
 has implications for the role of correlations in these: In tight-binding models of bilayer graphene, the dispersion is quadratic around the nodal points, yet the system develops a nematic instability at an infinitesimal interaction \cite{PhysRevB.81.041401}. For multiple-charge Weyl nodes and line-node semimetals with contact interaction, renormalization group theory indicates instabilities at a finite threshold \cite{PhysRevB.95.201102,PhysRevB.96.041113}. This is also true for single-layer graphene, where actual material parameters are situated close to a chiral symmetry-breaking regime \cite{PhysRevLett.111.056801}. Thus, semimetallic phases that depend explicitly on symmetry may be susceptible to correlation effects that destroy the underlying symmetry. 
 
 Reliably predicting the parameter regimes where symmetry-protected topological phases remain stable faces several delicate problems. The absence of screening means that interactions are effectively long-ranged, leading to infrared divergencies. Furthermore, competition between correlation effects originating from different length scales is very likely. For example, in graphene, the long-range part of the interaction drives the system towards an asymptotically free Dirac liquid with divergent Fermi velocity as $T\to 0$, without causing any instabilities \cite{PhysRevLett.118.026403}. Thus, the renormalization of the dispersion occurring primarily at the infrared end increases the effective kinetic energy relative to the short-range part of the interaction, which is believed to drive the phase transition. 
Therefore, an accurate solution to this class of problems requires that all length scales are treated on an equal footing. 
 
A type of symmetry-protected topological phase that has attracted considerable interest is the nodal-line semimetal. Predictions of this state has been made in TlTaSe$_2$ \cite{PhysRevB.93.121113}, CaAgP \cite{doi:10.7566/JPSJ.85.013708} and Ca$_3$P$_2$ \cite{,doi:10.1063/1.4926545} and ZrSiS \cite{Schoop2016} based on reflection symmetry. In PbTaSe$_2$ it has also been confirmed by angle resolved photo emission spectroscopy \cite{Bian2016}. 
Recently, the observations of strongly renormalized transport properties and Fermi velocity--as compared to DFT calculations--in ZrSiS was interpreted as an indication of a strongly correlated line-node semimetal \cite{Shao2020}.

In this work, we employ diagrammatic Monte Carlo simulations \cite{Van_Houcke_2010} to establish quantitative criteria for the stability of line-node semimetals in the presence of long-range interactions and also characterize the symmetry-broken phase which occurs for sufficiently strong coupling. We find evidence for a chiral insulator that supports metallic interfaces on domain walls that interpolate between different signs of the order parameter and can be manipulated in situ via an external field.

\section{Model}
We consider the case of a single nodal line with a bare Fermi velocity of $v_f^0$ running along the $z$-axis 
\bea
H_0(\bk)=v_f^0 \bk_{xy}\cdot\bsigma, \label{H0}
\eea
with an interaction of the form 
\bea
V(\bk)=\frac{\alpha}{\bk^2+\lambda^{-2}}.\label{interaction}
\eea
Here, $\lambda$ is a fictitious screening length introduced to regularize the series, and we are thus principally interested in the limit $\lambda\to \infty$. We consider a cylindrical domain given by 
\bea
|\bk_z|\le \Lambda,\; \sqrt{\bk_x^2+\bk_y^2}\le\Lambda, \label{domain}
\eea
where $\Lambda$ is the ultra violet cutoff. Because of a scale invariance associated with the linear dispersion, the only relevant length scale in the low-temperature limit is the ratio of the UV cutoff and the inverse screening length $\Lambda/\lambda^{-1}$. 
To see this, we may choose a temperature and energy scale where the temperature is unity by rewriting the partition function  $z(\beta,H)=z(1,\beta H)$. This gives a bare Greens function 
\bea
\frac{1}{i\omega -\beta H_0(\bk)}=G_0(\omega,\beta \bk), \;\omega=(2n+1)\pi, \label{bareGF}
\eea
where we have exploited the linearity of $H_0$ in $\bk$. Diagrammatic corrections to the Greens function take the form
\bea
\delta G(\omega,\beta \bk)=\prod_{i=1}^N d\bk_i \prod_{j=1}^{N}\beta V(\bk_j)\prod_{l=1}^{2N+1}G_0(\omega_l,\beta \bk_l),
\eea
where $N$ is the expansion order. If we introduce a change of scale $\bk'=\beta \bk$ we obtain
\bea
\delta G(\omega,\bk')= \prod_{i=1}^N  \frac{d\bk'_i}{\beta^D} \prod_{j=1}^{N}\beta V\Big[\frac{\bk'_j}{\beta}\Big]\prod_{l=1}^{2N+1}G_0(\omega_l, \bk'_l).
\eea
For a screened Coulomb interaction in $D=3$ we find 
\bea
\beta^{-D} \;\beta \frac{\alpha}{\bk'^2/\beta^2+\lambda^{-2}}=\frac{\alpha}{\bk'^2+\beta^2 \lambda^{-2}}.
\eea
The UV cutoff changes scales as $\Lambda\to \beta\Lambda$, giving 
\bea
G(\omega,\beta,\bk,\lambda^{-1},\Lambda)=G(\omega,1,\bk',\beta\lambda^{-1},\beta\Lambda), \label{scaleInvariance}
\eea
which is characterized by the ratio $\Lambda/\lambda^{-1}$ in the limit $\beta\to \infty$. 

In the perturbative regime, the nodal line (\ref{H0}) is protected by a symmetry due to being odd under an orthonormal map $\bk\to-\bk$. The implication of this symmetry is that on the $k_z$-axis, the Greens function must have a pole at zero energy as long as the series expansion remains convergent \cite{2018Symmetry}. Correspondingly, destroying the semimetallic phase requires breaking this symmetry. In a diagrammatic framework, this phase transition can be identified via a divergent susceptibility with respect to a symmetry-breaking perturbation.

\section{Contact interaction}
\begin{figure}[!htb]
	\includegraphics[width=\linewidth]{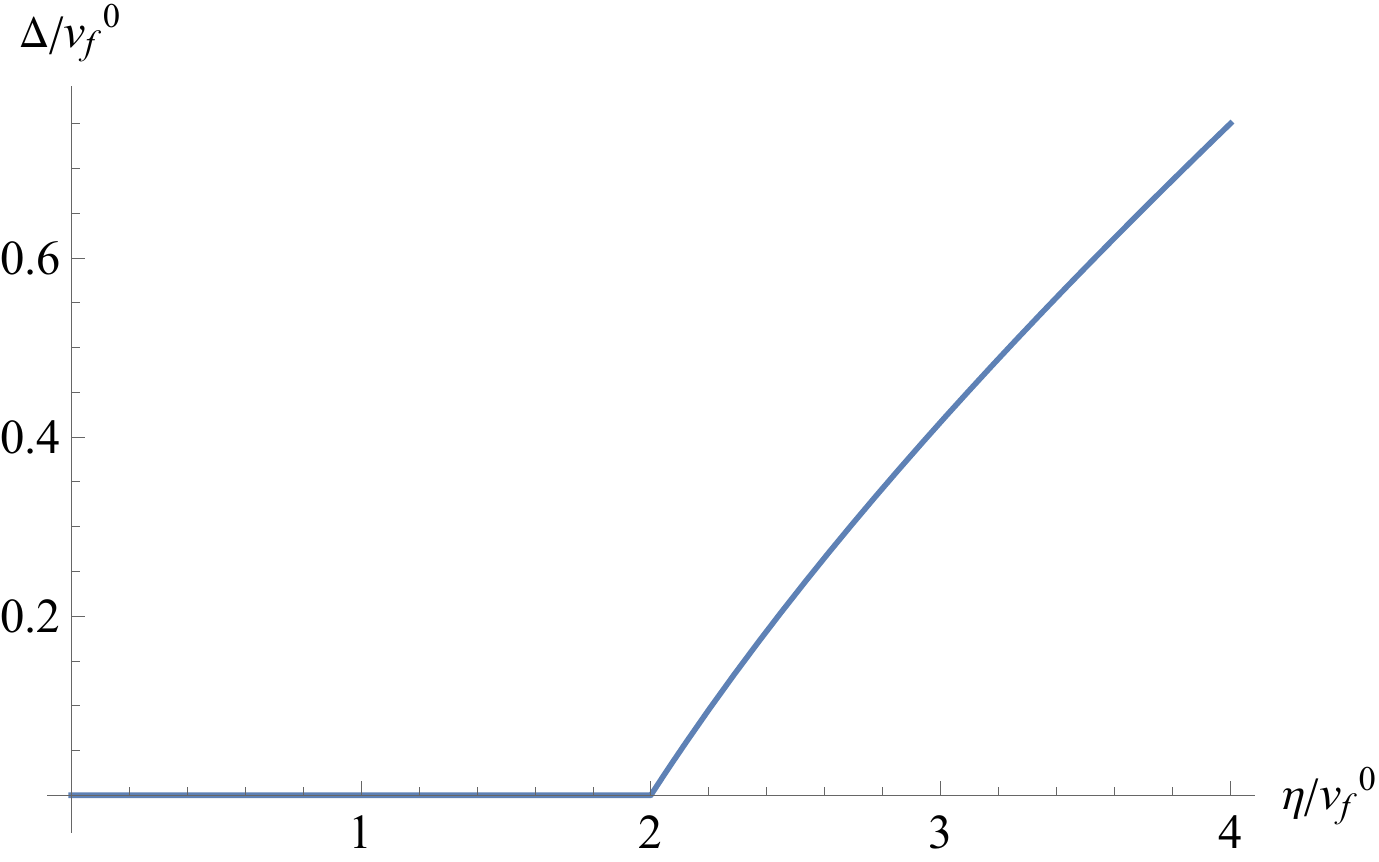}
	\caption{
		{\bf Self-consistent solution for the gap} with contact interactions, as a function of the rescaled coupling strength $\eta$. A second order transition is established at $\eta_c /v_f^0 =2$.
	}
	\label{contactcritical}
\end{figure}

For contact interaction, the self-consistent Fock theory can be solved analytically due to translation invariance in momentum space. 
Specifically, the self-energy satisfies the relation
\be \label{selfcE}
\Sigma(\omega_m,\bk) = \frac{1}{\beta} \sum\limits_{n} \int \frac{d^3q}{(2\pi)^3} V(\bq-\bk) \frac{1}{G_0^{-1}(\omega'_n,\bq) - \Sigma(\omega'_n,\bq) }.
\ee  
Here, it should be noted that at the level of Fock theory, $\Sigma$ is independent of frequency, and thus de facto takes the form of a correction to the effective dispersion. Furthermore, contact interaction does not renormalize the Fermi velocity since $H_0(\bk)$ is an odd function. Since the self energy is translation invariant, it must therefore take the form $\Sigma(\bk)= \Delta \sigma_z$.
Inserting this self-energy in (\ref{selfcE}) and summing over frequency, we obtain
\bea \nonumber
\Sigma(\bk) = \Delta \sigma_z = \int \frac{d^3q}{(2\pi)^3} \alpha \frac{H_0(\bq)+ \Sigma(\bq)}{2 \sqrt{(v_f^0 q_{xy})^2 + \Delta^2  } } \\
\times\tanh{ \frac{ \beta \sqrt{(v_f^0 q_{xy})^2 + \Delta^2  }   }{2}}, \label{selfeqdelta}
\eea 
where the integral of $H_0$ over $\bk$ vanishes. Thus, Eq. (\ref{selfeqdelta}) provides a self-consistent equation for $\Delta$ as a function of the coupling strength, whose solutions will provide the gap parameter in this regime. Solutions for which $\Delta$ is finite correspond to a symmetry-broken state, while the symmetric phase is characterized by a vanishing gap. 
In the low-temperature limit, and for a cylindrical domain (\ref{domain}) with $\Lambda=1$, the integral (\ref{selfeqdelta}) provides an algebraic expression for the gap of the form 
\be \label{selfsoldelta}
\frac{\eta}{2 v_f^0} \Bigg  (\sqrt{1+\frac{\Delta^2}{v_f^0}} - \left |\frac{\Delta}{v_f^0} \right|  \Bigg)-1 =0.
\ee
where we have introduced $\eta=\alpha(2\pi)^{-2}$. Equation (\ref{selfsoldelta}) predicts a critical coupling strength $\eta_c / v^0_f = 2$, see Fig. \ref{contactcritical}. Above this threshold, the gap is given by
\be
\frac{\Delta}{v_f^0} = \frac{\eta}{4 v_f^0} - \frac{v_f^0}{\eta}.
\ee
The onset of chiral a phase at a finite interaction strength is consistent with results from renormalization group theory for contact interaction \cite{PhysRevB.96.041113}.

\section{Simulations}
To check the stability of the semimetallic phase for long-range interactions, we employ diagrammatic Monte Carlo, which is a computational protocol based on the stochastic sampling of the diagrammatic expansion. 
Within this framework, the space of connected diagrams for the self-energy is sampled through a metropolis type random walk \cite{Van_Houcke_2010,PhysRevLett.119.045701,0295-5075-118-1-10004,PhysRevB.103.195147}. The Greens function is then obtained via Dyson's equation \cite{fetter}
\bea
G(\omega,\bk)=\frac{1}{i\omega-H_0(\bk)-\Sigma(\omega,\bk)}.
\eea
Here, we use a sampling protocol based on the worm algorithm as described in \cite{PhysRevB.97.075119}. We employ a bold scheme where the expansion is conducted in dressed Greens functions while retaining only skeleton graphs. Thus, at order $N=1$, the solution corresponds to self-consistent Fock theory. We do not employ bold interactions lines since this is expected to have little advantage for a semimetallic system. 

Following the scaling relation (\ref{scaleInvariance}) we can without loss of generality set $\Lambda=1$. This gives a volume of the momentum space of $2\pi$. We then rewrite the integral over $\bk$ as 
\bea
\int \frac{d\bk}{(2\pi)^D}=\frac{1}{(2\pi)}\int d\bk  \frac{\eta}{\alpha},\; \eta=\alpha (2\pi)^{-2},
\eea
which defines a rescaled interaction parameter $\eta$ and a set of units where the integral over momenta is of measure unity. 

We parameterize the temperature and scale in terms of a variable $\gamma$ so that
\bea \label{units}
\Lambda=1,\; \eta=2^{\gamma}\tilde{\eta}, \; v_f^0=2^{\gamma} \tilde{v}_f^0 , \;\lambda^{-1}=2^{-\gamma}\tilde{\lambda}^{-1}.\label{parameterization}
\eea
The limit $\gamma\to \infty$ thus corresponds to zero temperature and a divergent ratio  $\Lambda/\lambda^{-1}$. An observable that is convergent in this limit should correspondingly be a function of $\eta/v_f^0$.

To obtain a self-consistent solution for the model (\ref{H0}-\ref{interaction}), we consider a starting guess for the frequency-independent self-energy $\Sigma^0(\bk)$, which in turn provides a corresponding Greens function $G^0(\omega,\bk)$. A stochastic summation of the expansion in $V$ gives a new self-energy $\Sigma^1(\omega,\bk)$ which is subsequently used. This scheme is repeated until relevant observables have converged. Near the phase transition, this typically requires several hundred iterations. 
We have used two starting configurations for the self-energy, featuring extremely small or relatively large symmetry-breaking terms, respectively. For most parameter regimes, these result in identical solutions. However, at low temperatures and for a coupling strength that is slightly larger than the critical coupling, we observe a family of very fragile meta-stable symmetric solutions that likely result from competition between the symmetric and antisymmetric parts of the self-energy. 

To track the onset of a symmetry-broken phase, we define the chiral order parameter as follows. First, we note that the frequency-independent part of the self-energy can be written 
\bea
\Sigma(\bk)=\bf{d}({\bk})\cdot\bsigma.
\eea
The chiral symmetry-breaking is generated by the $z$-component, prompting us to define an order parameter of the form
\bea
O=\frac{1}{{2\pi}}\int d\bk {\bf d}_z (\bk).
\eea
Since we consider a straight nodal line, we assume a solution that is translation invariant in the $z-$direction. Furthermore, we assume a symmetry of the self-energy
\bea
e^{-i \phi \sigma_z/2} \Sigma(\omega,\bk) e^{i \phi \sigma_z/2}= \Sigma(\omega,R^z_\phi\bk)
\eea
where $R^z_\phi$ represents a rotation around the $z-$axis by $\phi$. 

\begin{figure}[!htb]
\includegraphics[width=\linewidth]{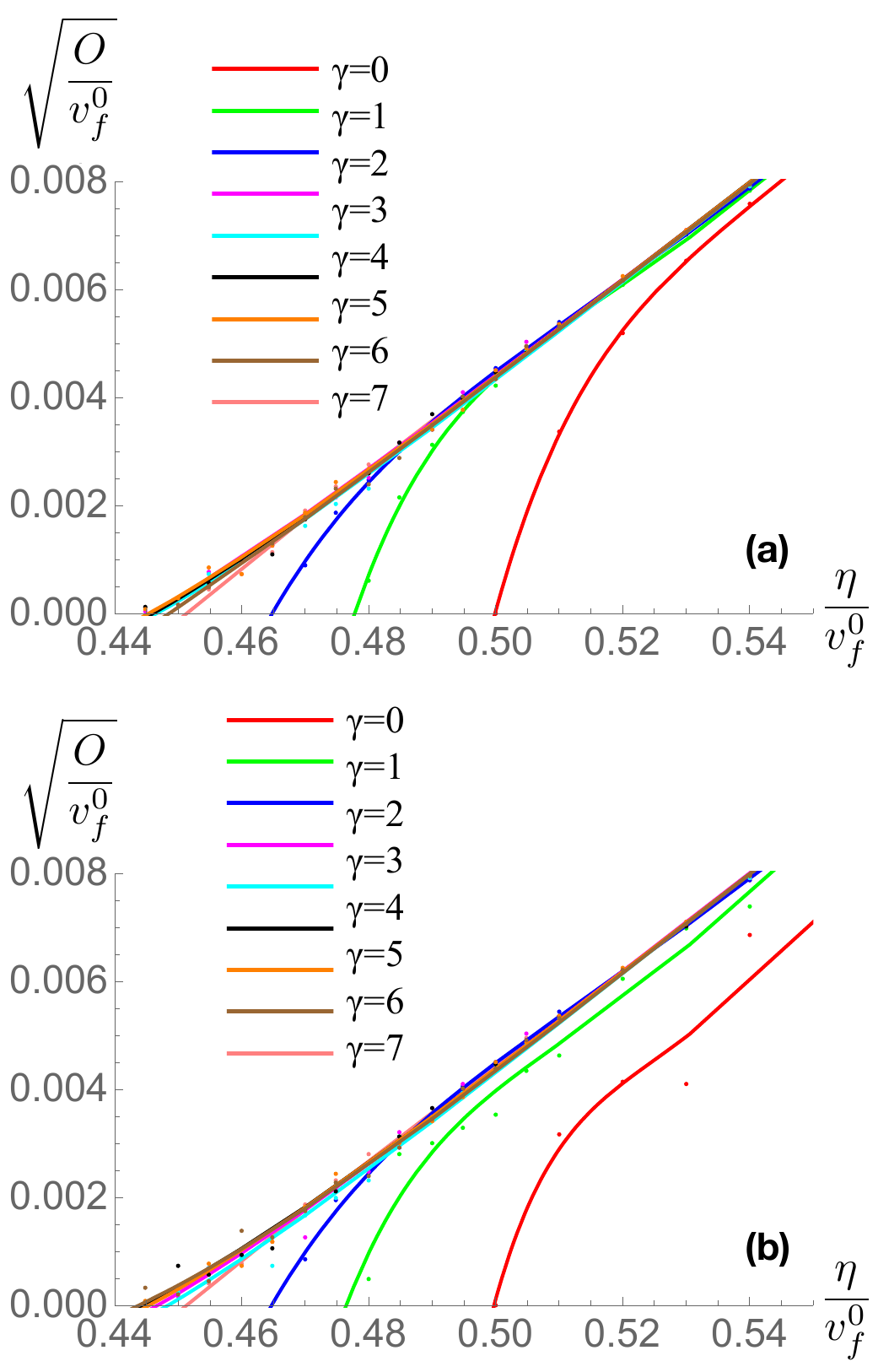}
\caption{
{\bf The chiral order parameter} $O$ as a function of the interaction strength $\eta$ in units of the bare Fermi velocity for Fock theory (a), and second-order theory (b). Note that we display the square root of the order parameter since it is approximately quadratic in $\eta$ over most parameter ranges. Here, the screening parameter is $\lambda^{-1}=10^{-2}\times 2^{-\gamma}$, while $\beta v_f^0=10^2 \times 2^{\gamma}$ so that the solutions correspond to progressively lower temperatures and longer screening lengths. 
For larger values of $\gamma$, the solutions collapse onto a single line indicating that the chiral order remains convergent in the infrared limit.
}
\label{order}
\end{figure}

The results from the diagrammatic Monte Carlo simulations are summarized in Fig. \ref{order}. For most parameter ranges, the order parameter scales approximately as $O\sim (\eta/v_f^0)^2$, prompting us to plot the square root. The solutions correspond to different values of $\gamma$, which controls the model parameters according to (\ref{parameterization}).
As we progressively decrease the temperature and increase the screening length, the order parameter saturates to a single line which depends only on  $\eta/v_f^0$, indicating that the chiral order exhibits a well-defined IR limit at zero temperature. The presented data corresponds to a first and second-order expansion. 

 \begin{figure}[!htb]
\includegraphics[width=\linewidth]{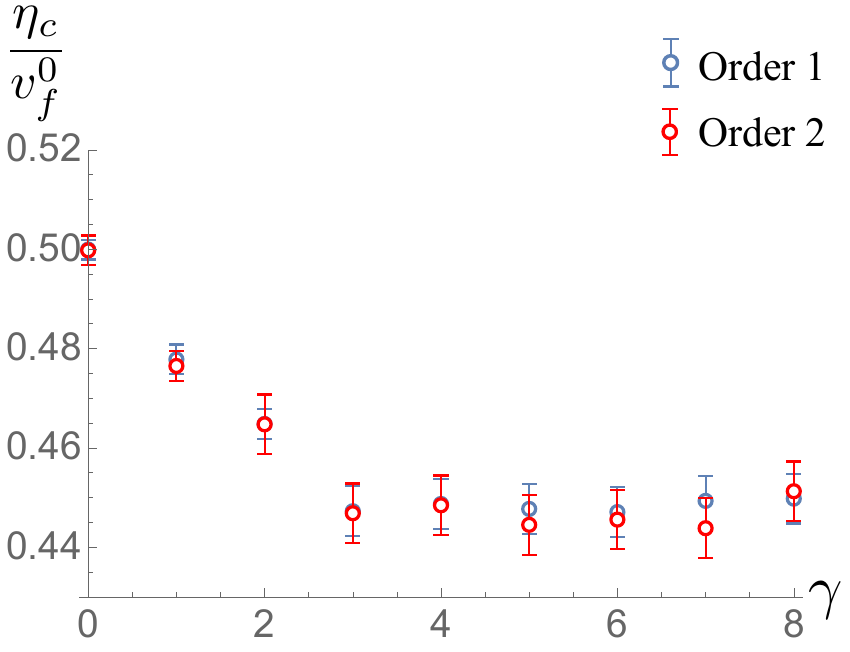}
\caption{
{\bf Critical coupling strength} int units of the bare Fermi velocity $v_f^0$ for progressively lower temperatures and longer screening lengths, parameterized as $\lambda^{-1}=10^{-2}\times 2^{-\gamma}$, and $\beta v_f^0=10^2 \times 2^{\gamma}$. For $\gamma\ge 3$ we estimate the critical coupling to $\eta_c/v_f^0=0.45\pm 0.01$. The corrections at second order are not discernible at this accuracy. 
}
\label{fit}
\end{figure}

In Fig. \ref{fit} we see the critical coupling strength as a function of $\gamma$, extracted from the data presented in Fig. \ref{order}. The critical point saturates to $\eta_c/v_f^0=0.45\pm 0.1$. The correction from first to second order falls within the error bars, indicating that this problem is well captured by self-consistent Fock theory. 
This is consistent with previous applications of diagrammatic techniques to semimetallic systems: In Weyl semimetals, the correction to the Greens function is almost entirely contained in the frequency-independent part of the self-energy, leading to the emergence of virtually free fermions \cite{PhysRevB.98.241102}. In graphene, at least the long-range part of the interaction drives the system towards an asymptotically free Dirac liquid \cite{PhysRevLett.118.026403}, while for short-range interactions, the convergence of the series has been demonstrated analytically up to a finite threshold \cite{PhysRevB.79.201403}. Most likely, this results from the exponential suppression of diagram topologies that involve excitation of the background in semimetals.

\section{Landau levels and magnetic response}
To compute the Landau levels arising when the line-node is placed in a magnetic field, we consider a dispersion of the form (\ref{H0}) and take the Fermi velocity to be unity. This gives 
\be \label{ModelH}
H(\bk) = k_x \sigma_x + k_y \sigma_y = \bpm 0 & k_x - i k_y \\ k_x + i k_y & 0 \epm,
\ee
with energy bands
\be
\varepsilon_\pm = \pm \sqrt{k_x^2 + k_y^2}.
\ee
Thus, the nodal line runs along the $z$-axis $(0,0,k_z)$.

To couple the system to an external magnetic field, we introduce the displacement of the momentum $\bk$ by the vector potential $\bA$:
\be
\bk \rightarrow \bk' = \bk + \bA.
\ee

First, we consider a magnetic field along the $z$-direction, i.e., $\bB = B \bs{\hat z}$. Working in the axial gauge, the vector potential reads $\bA = (-B y/2, B x/2, 0)$, so that we can define the ladder operators in terms of the new momenta
\be \label{Ladder-Bz}
a = \frac{k_x'-i k_y'}{\sqrt{2 B}}, \qquad \qquad a^\dagger = \frac{k_x'+i k_y'}{\sqrt{2 B}},
\ee
which allows us to write the Hamiltonian in the form
\be
H = \bpm 0 & \sqrt{2 B} a \\ \sqrt{2 B} a^\dagger & 0 \epm.
\ee

The Landau levels can be easily found from the eigenequation $H \Phi = E \Phi$. For $\Phi = ( |\phi_1 \rangle, \,  |\phi_2 \rangle)^{\rm T}$, we obtain the two equations 
\bea
\sqrt{2 B} a |\phi_2 \rangle &=& E |\phi_1 \rangle,  \\
\sqrt{2 B} a^\dagger |\phi_1 \rangle &=& E |\phi_2 \rangle, 
\eea
and, by inserting the first equation into the second, we arrive at
\be
2 B a^\dagger a | \phi_2 \rangle = E^2 | \phi_2 \rangle,
\ee
which describes a harmonic oscillator with 
\be
E = \pm \sqrt{2 B n}, \qquad |\phi_2 \rangle = c \, |n \rangle, \qquad n = 0, 1, 2, \dots \in \mathbb N,
\ee
where $c$ is a normalization constant. Then, for $| \phi_1 \rangle$ we have
\be
|\phi_1 \rangle = \pm c \, |n-1 \rangle.
\ee
Therefore, for a magnetic field in the $z$ direction, the Landau levels are given by
\be \label{Eq-LLz}
E_\pm = \pm \sqrt{2 B n},
\ee
with eigenstates
\be
\Phi_\pm = c \bpm \pm |n-1 \rangle \\ |n \rangle \epm.
\ee

\begin{figure}[t]
\begin{center}
\includegraphics[width=\linewidth]{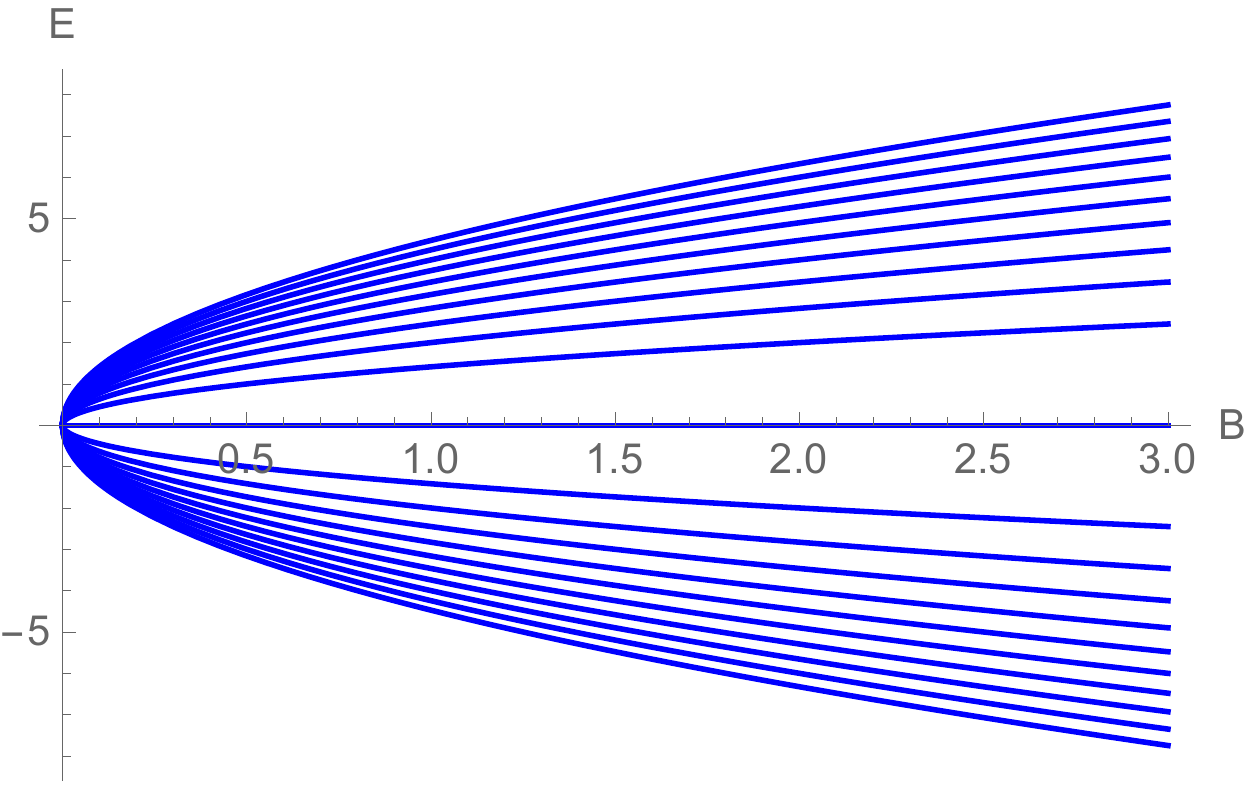}
\end{center}
\caption{Landau levels for the line-node semimetal in the presence of an external field $\bB = B \bs {\hat z}$, as a function of $B$ as given by Eq. (\ref{Eq-LLz}). Since the dispersion is independent of $k_z$, the energy levels are flat in all directions. A zero-energy mode remains for any value of the applied field.}
\label{Fig-LLz}
\end{figure}
 The eigenstates appear as a spectrum of bands that are flat in all directions. The gap between these is controlled by the external field, as shown in Fig. \ref{Fig-LLz}.
In contrast to Weyl semimetals, line nodes do not give rise to a chiral anomaly \cite{RevModPhys.90.015001} in the presence of a magnetic field. Instead, a single band remains at the Fermi level because the dispersion is independent of $k_z$.   

For an in-plane field of the form $\bB = B \bs{\hat x}$, we may choose a vector potential of the form $\bA = (0, -B z/2, B y/2)$.
The ladder operators can then be constructed as 
\be \label{Ladder-Bx}
a = \frac{k_y'-i k_z'}{\sqrt{2 B}}, \qquad \qquad a^\dagger = \frac{k_y' + i k_z'}{\sqrt{2 B}}.
\ee
Expressed in this language, the Hamiltonian (\ref{ModelH}) takes the form
\be
H =  \bpm 0 & k_x - i \sqrt \frac{B}{2} (a + a^\dagger) \\  k_x + i \sqrt \frac{B}{2} (a + a^\dagger) & 0 \epm.
\ee
From the eigenequation $H \Phi = E \Phi$, we obtain for $|\phi_1 \rangle$ and $|\phi_2 \rangle$
\bea 
\left[ k_x - i \sqrt \frac{B}{2} (a + a^\dagger) \right] |\phi_2 \rangle &=& E |\phi_1 \rangle,  \\
\left[ k_x + i \sqrt \frac{B}{2} (a + a^\dagger) \right] |\phi_1 \rangle &=& E |\phi_2 \rangle.
\eea
As before, we can take the first equation and plug it into the second. By using the commutator $[a, a^\dagger] = 1$, we finally arrive at
\be 
\left[k_x^2 + \frac{B}{2} \left(a^2 + (a^\dagger)^2 + 2 a^\dagger a + 1\right) \right] |\phi_2 \rangle = E^2 |\phi_2 \rangle. \label{phi2Eq}
\ee
Solving the equation (\ref{phi2Eq}) is complicated by the presence of terms of the form
$\sim a^2$ and $\sim( a^\dagger)^2$, which render it anharmonic so that standard recipes for extracting the Landau levels are not applicable.   
For this reason, we adopt the Bargmann representation \cite{Bargmann1962, Bargmann1961,Bargmann1967}, which has been widely used in this scenario. Notably, this technique was applied to an anharmonic oscillator with a quartic potential \cite{doi:10.1063/1.522747} and the two-mode squeeze harmonic oscillator and the $k$th-order harmonic generation \cite{Zhang2013}. In this representation, the ladder operators are related to a complex variable $z$ according to
\be
a^\dagger = z, \qquad \qquad a = \frac{d}{dz},
\ee
whilst the wave function is a holomorphic function of $z$ only, namely, 
\be
|\phi_1 \rangle = \varphi_1 (z), \qquad \qquad  |\phi_2 \rangle = \varphi_2 (z).
\ee
Expressed in this formalism, Eq. (\ref{phi2Eq}) takes the form  
\be
\frac{B}{2} \varphi_2'' + B z \varphi' + \left[ \frac{B}{2} (z^2+1) + k_x^2 \right] \varphi_2 = E^2 \varphi_2,
\ee
where we have used the notation $\varphi_2' = \frac{d \varphi_2}{d z}$ and $\varphi_2'' = \frac{d^2 \varphi_2}{d z^2}$.

\begin{figure}[t]
\begin{center}
\includegraphics[width=\linewidth]{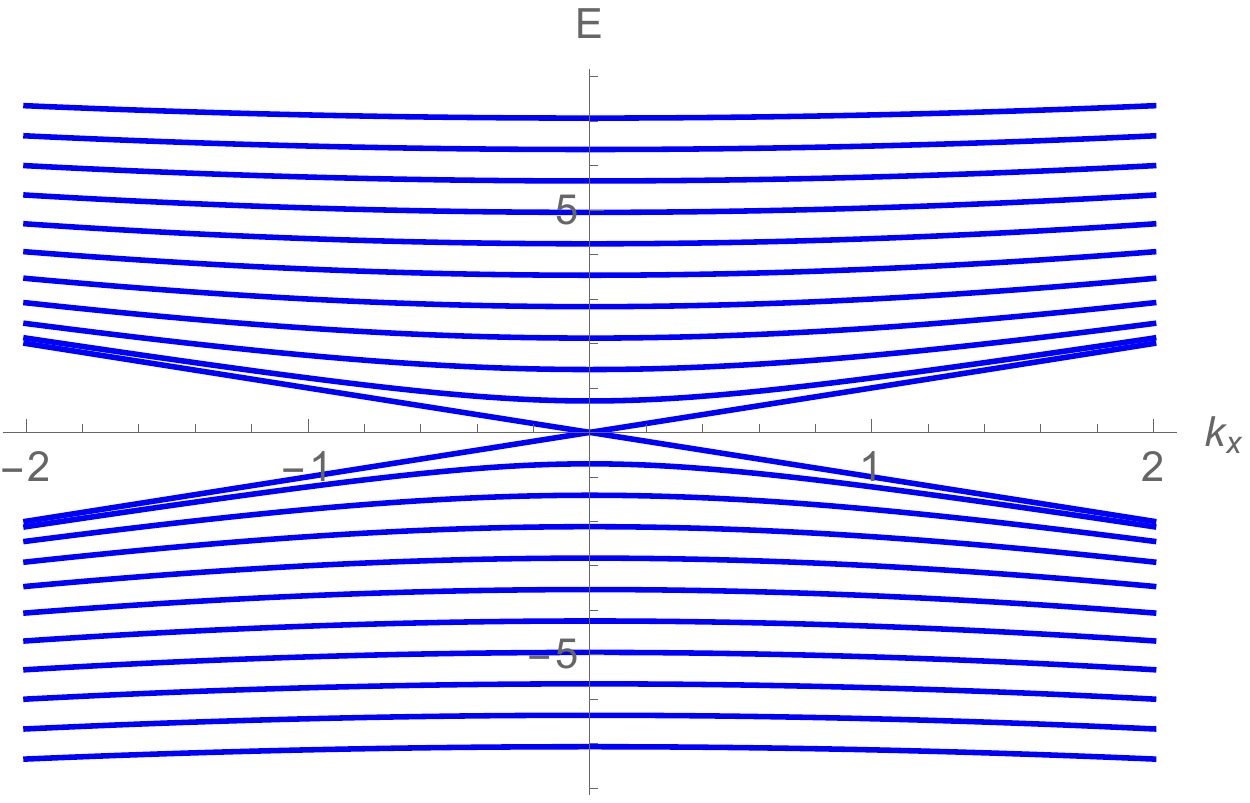}
\end{center}
\caption{Landau levels for the line-node semimetal in the presence of an external field $\bB = \bs{\hat x}$ ($B = 1$) as given by Eq. (\ref{Eq-LLx}). In contrast to the case of a magnetic field along the $\bs{\hat z}$ direction in Fig. \ref{Fig-LLz}, the levels do exhibit a momentum dependence. The two energy levels that are closest to the Fermi surface cross in the origin. }
\label{Fig-LLx}
\end{figure}

The different quantum states correspond to solutions of this equation for corresponding quantum numbers, such as polynomials of degree $n$. 
In principle, it is possible to extract a solution in the form of a power series in $z$, though this turns out to be highly inefficient. Thus, we instead introduce a reparameterization of $\varphi_2 (z)$ given by
\be
\varphi_2 (z) = e^{-z^2/2} \, \psi_2 (z).
\ee
The differential equation then takes the form
\be
\psi_2'' + \omega^2 \, \psi_2 = 0, \qquad {\rm with} \quad \omega^2 = \frac{2}{B} (k_x^2-E^2).
\ee
The trivial solution consisting of a combination of two exponentials does not correspond to the Landau levels, and the energy $E$ still appears as an arbitrary constant. To extract the nontrivial solutions, we need to introduce a change of variables of the form
\be
\rho = e^z  \quad \Rightarrow \quad z = \ln \rho
\ee
which finally transforms the equation for $\psi_2$ into
\be \label{psi2Eq}
\rho^2 \frac{d^2 \psi_2}{d \rho^2} + \rho \frac{d \psi_2}{d \rho} + \omega^2 \, \psi_2 = 0.
\ee
Now, we may find solutions of this equation as a polynomial in $\rho$ of degree $n$ by considering
\be
\psi_2(\rho) = \sum_{i=0}^n f_i \, \rho^i.
\ee
Plugging this into the equation we obtain an expression in terms of the coefficients of the expansion $f_i$
\be
\omega^2 f_0 + (1+\omega^2) f_1 \, \rho + \sum_{i=2}^n (i^2 + \omega^2) f_i \, \rho^i = 0.
\ee
Since our assumption of a polynomial of degree $n$ implies $f_n \neq 0$, it trivially follows that $n^2+\omega^2 =0$, giving Landau levels with an energy
\be \label{Eq-LLx}
E_\pm = \pm \sqrt{\frac{1}{2} n^2 B + k_x^2},
\ee
together with $f_i = 0, \; \forall \, i \neq n$.
The function $\psi_2$ then reads
\be
\psi_2(\rho) = c \, \rho^n \quad \Rightarrow \quad \psi_2 (z) = c \, e^{n z},
\ee
where $c$ is a normalization constant. Returning to the original wave function component $\varphi_2$ we get
\be
\varphi_2 (z) = c \, e^{n z - z^2/2}.
\ee
The $\varphi_1$ component may be written as
\be
\varphi_{1, \pm} (z) = \frac{c}{E_\pm} \left(k_x - i \sqrt \frac{B}{2} n \right) \, e^{n z - z^2/2}.
\ee
In contrast to the Weyl semimetals, the states corresponding to $n=0$ can be treated on an equal footing with the remaining levels. This quantum number gives two different states related by $E_1=-E_2$ that cross the Fermi level at $k_x=0$, as depicted in Fig. \ref{Fig-LLx}. 

Finally, we will review these results once a symmetry breaking term of the form $\Delta \, \sigma_z$ is included in the Hamiltonian. This gives 
\be
H(\bk) = k_x \sigma_x + k_y \sigma_y + \Delta \, \sigma_z = \bpm \Delta & k_x - i k_y \\ k_x + i k_y & -\Delta \epm.
\ee
Hence, the energy bands are now given by
\be
\varepsilon_\pm = \pm \sqrt{k_x^2 + k_y^2 + \Delta^2},
\ee
so that the system is an insulator. The symmetry-breaking perturbation introduced in the system has now gapped out the nodal line, suggesting that the Landau levels will form away from the Fermi level irrespectively of the orientation of the magnetic field.

For the case of an external field $\bB = B \bs{\hat z}$ with ladder operators defined according to Eq. (\ref{Ladder-Bz}), the Hamiltonian reads
\be
H = \bpm \Delta & \sqrt{2 B} a \\ \sqrt{2 B} a^\dagger & -\Delta \epm.
\ee
Following the same approach as above and considering the eigenequation $H \Phi = E \Phi$, we obtain a harmonic oscillator-like equation for $|\phi_2 \rangle$ with the solution
\be
E_\pm = \pm \sqrt{2 B n + \Delta^2}, \quad |\phi_2 \rangle = c \, |n \rangle,
\ee
where $c$ is a normalization. Comparing with Eq. (\ref{Eq-LLz}), we see that the effect of the perturbation $\Delta$ is to introduce a displacement of the Landau levels. On the other hand, for $|\phi_1 \rangle$ we have
\be
|\phi_{1, \pm} \rangle = c \, \frac{\sqrt{2 B n}}{E_\pm - \Delta} |n-1 \rangle.
\ee

\begin{figure}[t]
\begin{center}
\includegraphics[width=\linewidth]{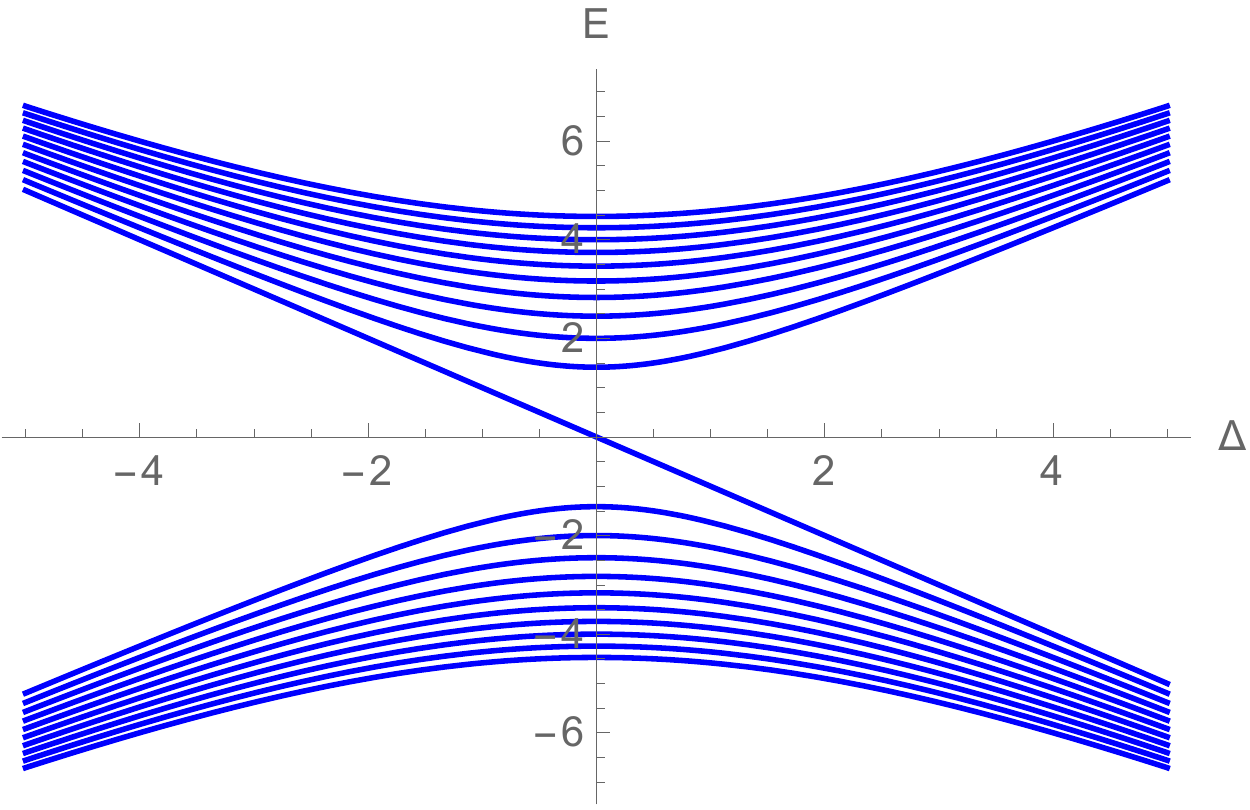}
\end{center}
\caption{Landau levels for a symmetry-broken line-node semimetal in the presence of an external field $\bB = \bs{\hat z}$ ($B = 1$) as a function of the symmetry breaking parameter $\Delta$. 
The applied field explicitly breaks the symmetry between the chiralities $\Delta$ and $-\Delta$ respectively, indicating that the chiral order can be manipulated via a magnetic field.
}
\label{Fig-epsLLz}
\end{figure}

This scenario is slightly different from the unperturbed case, and the value $n=0$ needs to be considered separately since $|\phi_1 \rangle = 0$ and $|\phi_2 \rangle = |0 \rangle$. As a result, the energy for $n = 0$ is given by $E_0 = - \Delta$. In conclusion, we thus find
\be
E_0 = - \Delta, \qquad \qquad E_{n, \pm} = \pm \sqrt{2 B n + \Delta^2}, \quad \forall \; n \neq 0.\label{Bchiral}
\ee
Fig. \ref{Fig-epsLLz} shows the Landau levels (\ref{Bchiral}) for $B=1$ as a function of the symmetry breaking parameter $\Delta$. Besides the aforementioned displacement of the energy levels for $n \neq 0$, the unperturbed $E_0 = 0$ state is shifted relative to the Fermi level depending on $\Delta$ in a similar manner to the $k_z$-dependence of the Landau levels appearing in Weyl semimetals. 
Thus, for a symmetry-broken state, the magnetic field lifts the degeneracy between the two chiralities, implying that this degree of freedom can be manipulated by an external field in experiments. 

Finally, we consider the case of an in-plane magnetic field $\bB = B \bs{\hat x}$ in the symmetry-broken phase. Expressed in the ladder operators defined in Eq. (\ref{Ladder-Bx}), the Hamiltonian takes the form
\be
H = \bpm \Delta & k_x - i \sqrt \frac{B}{2} (a + a^\dagger) \\  k_x + i \sqrt \frac{B}{2} (a + a^\dagger) & -\Delta \epm.
\ee
As before, the eigenvalue equation gives rise to an anharmonic problem, meaning that we have to rely on the Bargmann representation. For $|\phi_2 \rangle$ the solution is given by Eq. (\ref{phi2Eq}), except for a shift in energy given by $E^2 \rightarrow E^2 - \Delta^2$. Hence, we can use the same protocol as above for the in-plane field with the $\omega$ parameter accordingly modified to
\be
\omega^2 = \frac{2}{B} (k_x^2 - E^2 + \Delta^2).
\ee
The Landau levels are given by
\be
E_\pm = \pm \sqrt{\frac{1}{2} n^2 B + k_x^2 +\Delta^2},
\ee
with the wave function components
\bea
\phi_{1, \pm} (z) &=& \frac{c}{E_\pm - \Delta} \left(k_x - i \sqrt \frac{B}{2} n \right) e^{n z - z^2/2},  \\
\phi_2 (z) &=& c \, e^{n z - z^2/2},
\eea
where $c$ is a normalization constant.

\begin{figure}[t]
\begin{center}
\includegraphics[width=\linewidth]{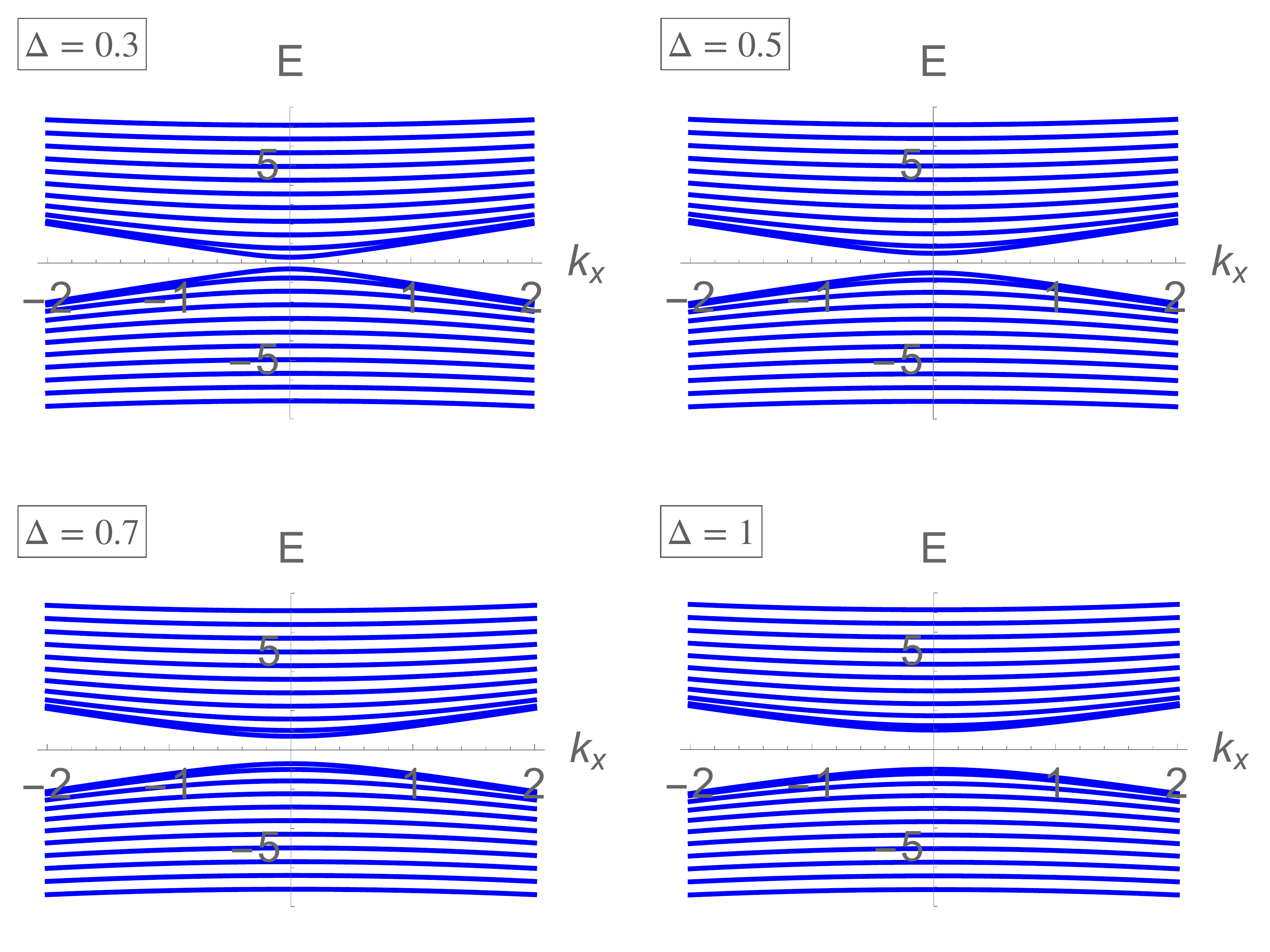}
\end{center}
\caption{Landau levels for the symmetry-broken line-node semimetal in the presence of an external field $\bB = \bs{\hat x}$ ($B = 1$). As the system becomes gapped, the crossing of the Landau levels shown in Fig. \ref{Fig-LLx} is lifted.}
\label{Fig-epsLLx}
\end{figure}
In the symmetry-broken phase, the system becomes gapped, with Landau levels situated at finite energies, as seen in Fig. \ref{Fig-epsLLx}. The levels closest to the Fermi surface attain a gap of $|\Delta|$ which is thus independent of the chirality.

\section{Interface states}
In the symmetric phase, the line-node semimetals exhibit metallic drum-head surface states \cite{PhysRevB.93.121113}, which are generalizations of the Fermi arcs that occur in the Weyl semimetals \cite{Jia2016}. These states are stabilized by a combination of topology and symmetry in the sense that the states are topologically protected in a subspace generated by the symmetry. Once this symmetry is spontaneously broken, the lines are gapped out, and the edge states are no longer protected. 

However, the symmetry-broken phase permits domain walls that interpolate between regions of different chirality, on which the symmetry-breaking term $\Delta$ changes sign.  
 This opens the possibility for metallic interface states that are bound to these topological defects. To model this scenario, we consider a domain wall described by %
\be
\Delta = \Delta(y) = \left\{ \begin{array}{cc} \Delta_+, & y>0 \\ \Delta_- & y<0  \\ 0 & y=0  \end{array} \right. ,\label{DeltaC}
\ee
where $\Delta_+>0$ and $\Delta_- <0$ are constants.

To identify the interface states, we apply an analytical approach based on trial functions that has been applied to Fermi arcs within Weyl semimetals in semi-infinite systems \cite{Zhang2016,Ojanen2013}. Since $\Delta(y)$ is translation invariant in the $\bs{\hat x}$ and $\bs{\hat z}$ directions but not along $\bs{\hat y}$, it follows that $k_x$ and $k_z$ are good quantum numbers while $k_y$ is not. Hence, we conduct the substitution $k_y \rightarrow - i \partial_y$, which transforms the perturbed Hamiltonian into
\be
H (k_x , - i \partial_y, k_z,y) = k_x \sigma_x - i \partial_y \sigma_y + \Delta(y) \sigma_z.
\ee

Next, we introduce a trial wave function of the form
\be \label{Trial-wf}
\Psi(x,y,z) = \psi_\lambda |x,z \rangle = \bpm \psi_1\\ \psi_2 \epm e^{\lambda y} |x,z\rangle.
\ee
Therefore, our problem is reduced to the eigenequation
\be
H (k_x , - i \partial_y, k_z,y) \Psi = E \Psi,
\ee
with a continuity condition at $\Psi(y=0)$. The secular equation, ${\rm det} |H (k_x , - i \partial_y, k_z,y) -E| = 0$, may be used to find the possible values of $\lambda$, namely,
\be
\lambda = \pm \sqrt{k_x^2 + \Delta^2 - E^2}.
\ee
Requiring the wave function to vanish at $y \rightarrow \pm \infty$, we need to separate the two regions of different chirality into $\Psi_+$ for $y>0$ and $\Psi_-$ for $y<0$. Then, we obtain
\bea
\Psi_+ &=& c_+ \psi_{\lambda_+} |x,z \rangle = c_+ \bpm \psi_1^+ \\ \psi_2^+ \epm e^{\lambda_+ y} |x,z \rangle, \\
\nonumber \\
\Psi_- &=& c_- \psi_{\lambda_-} |x,z \rangle = c_- \bpm \psi_1^- \\ \psi_2^- \epm e^{\lambda_- y} |x,z \rangle,
\eea
where $c_\pm$ are constants whilst
\be
\lambda_\pm = \mp \sqrt{k_x^2 +\Delta_\pm^2 - E^2}.
\ee

On the other hand, for the eigenstates, there are two possible sets of spinors $\psi^\pm = (\psi_1^\pm, \, \psi_2^\pm)^{\rm T}$:
\be \label{SetSpinors}
\psi^\pm = \bpm \lambda_\pm - k_x \\ \Delta_\pm - E \epm \qquad {\rm and} \qquad \psi^\pm = \bpm \Delta_\pm + E \\ \lambda_\pm + k_x \epm.
\ee
Imposing that the solution is continuous at $y=0$ we obtain 
\be
c_+ \psi_{\lambda_+} (y=0,E) = c_- \psi_{\lambda_-} (y=0,E),
\ee
or equivalently,
\be
c_+ \psi^+ - c_- \psi^- = 0. \label{match}
\ee

Thus, the condition (\ref{match}) gives us a system of two equations with two unknowns, $c_+$ and $c_-$. Hence, to have a nontrivial solution, it is necessary that
\be
{\rm det} |\psi^+ \quad - \psi^-| = 0.
\ee
Imposing this condition on the eigenvectors defined by Eq. (\ref{SetSpinors}) and using Eq. (\ref{DeltaC}) we find 
\be
-(\lambda_+ - k_x) E + (\lambda_- - k_x) E = 0,
\ee
\be
(\lambda_- + k_x) E - (\lambda_+ + k_x) E = 0,
\ee
which reduces to
\be
\lambda_+ = \lambda_- = 0 \qquad \Rightarrow \qquad E = \pm k_x
\ee

Therefore, we obtain two localized states proximate to the interface $y=0$ given by the wave functions
\be
\Psi_\pm (E = -k_x) = c_\pm \bpm -\Delta_\pm - k_x \\ \Delta_\pm + k_x \epm e^{- \Delta_\pm y} \, |x, z \rangle,
\ee
\be
\Psi_\pm (E = k_x) = c_\pm \bpm \Delta_\pm + k_x \\ -\Delta_\pm + k_x \epm e^{- \Delta_\pm y} \, |x, z \rangle,
\ee
where  $c_\pm$ is the normalization. At $k_x=0$ these meet at the Fermi level, implying a metallic interface state in the form of a  line-node that is exponentially localized to the domain wall. 

To solve the problem of interface states on a domain wall for a more realistic gap $\Delta$ which is continuous in $y$, it is generally necessary to apply numerical methods since the analytical technique introduced above cannot be applied when the gap has an explicit dependence on $y$.
 To obtain a numerically tractable problem in this scenario, we first conduct an inverse Fourier transform on $y$ and consider a finite system that can be diagonalized to find the possible surface states.
For this purpose, we consider a Hamiltonian which is periodic in $k_y$ instead of its continuum equivalent:
\be
H = \sin k_x \sigma_x + \sin k_y \sigma_y.
\ee
Explicitly writing the creation and annihilation operators, we find
\be
H = \sum_\bk \left[ \sin k_x (a_\bk^\dagger b_\bk + b_\bk^\dagger a_\bk) + i \sin k_y (- a_\bk^\dagger b_\bk + b_\bk^\dagger a_\bk) \right].
\ee
The inverse Fourier transforms along the $\bs{\hat y}$ direction takes the form
\be
a_\bk = \frac{1}{\sqrt M} \sum_j e^{-i k_y j} a_{\bk_\parallel, \,j}. \qquad b_\bk = \frac{1}{\sqrt M} \sum_j e^{-i k_y j} b_{\bk_\parallel, \, j},
\ee
where $M$ corresponds to the number of layers in the $\bs{\hat y}$ direction, $j$ is the layer index, and $\bs{\bk_\parallel}$ denotes the momentum parallel to the (010) surface. 
Thus, we obtain
\bea
H = \sum_{\bk_\parallel,  j} \Big[\sin k_x \bs{c}_{\bk_\parallel,  j}^\dagger  \sigma_x  \bs{c}_{\bk_\parallel,  j} - \frac{i}{2}  \bs{c}_{\bk_\parallel,  j}^\dagger  \sigma_y   \bs{c}_{\bk_\parallel,  j+1} \\+ \frac{i}{2}  \bs{c}_{\bk_\parallel,  j+1}^\dagger \, \sigma_y   \bs{c}_{\bk_\parallel,  j} \Big],
\eea
where we have defined
\be
 \bs{c}_{\bk_\parallel, \, j} = (a_{\bk_\parallel, \, j} , \, b_{\bk_\parallel, \,  j})^{\rm T}.
\ee

\begin{figure}[t]
\begin{center}
\includegraphics[width=\linewidth]{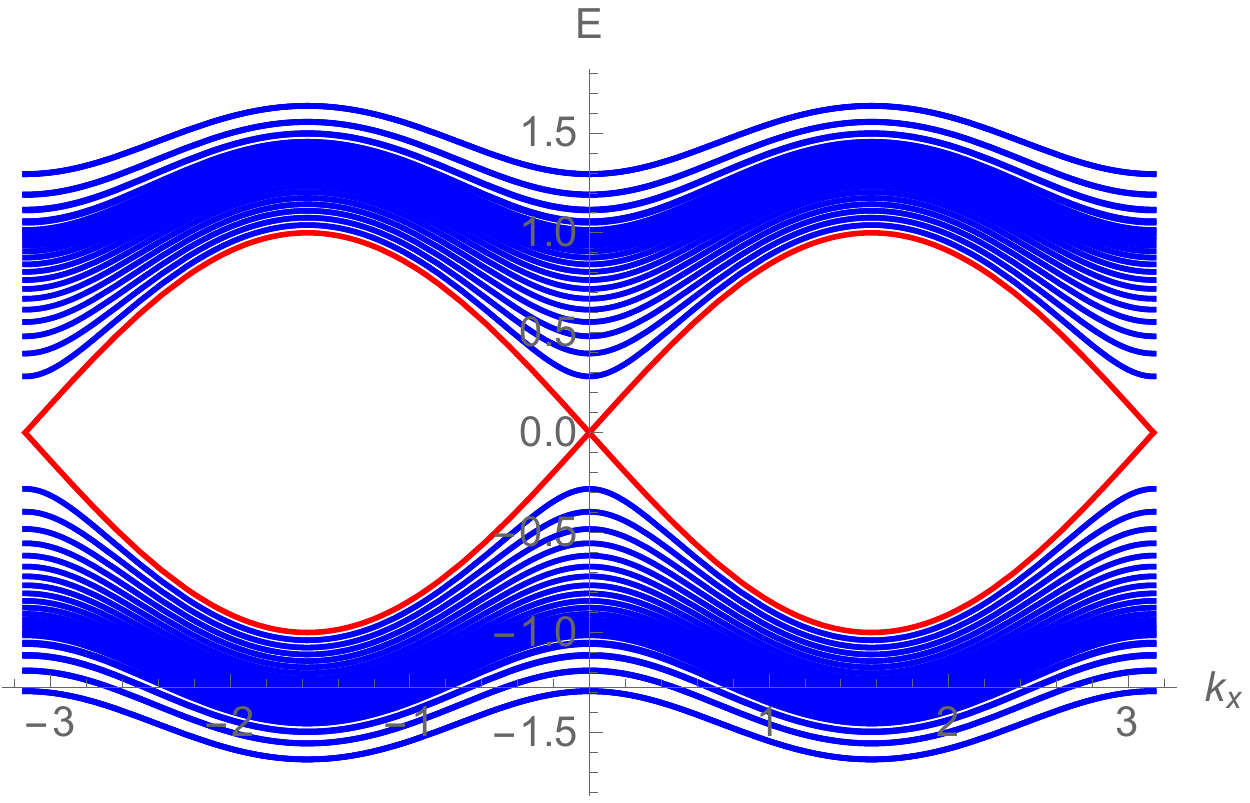}
\end{center}
\caption{
{\bf Metallic interface states} situated at a domain wall that interpolates between different signs on the chiral order parameter. The energy levels correspond to a system with $51$ layers in the $\bs{\hat y}$-direction with a linear symmetry-breaking term given by Eq. (\ref{Linear-Delta}). 
The red lines correspond to a family of solutions that are exponentially localized to the middle layer that exhibits a line node at $k_x=0$.}
\label{Fig-EnergyBands}
\end{figure}

\begin{figure}[t]
\begin{center}
\includegraphics[width=\linewidth]{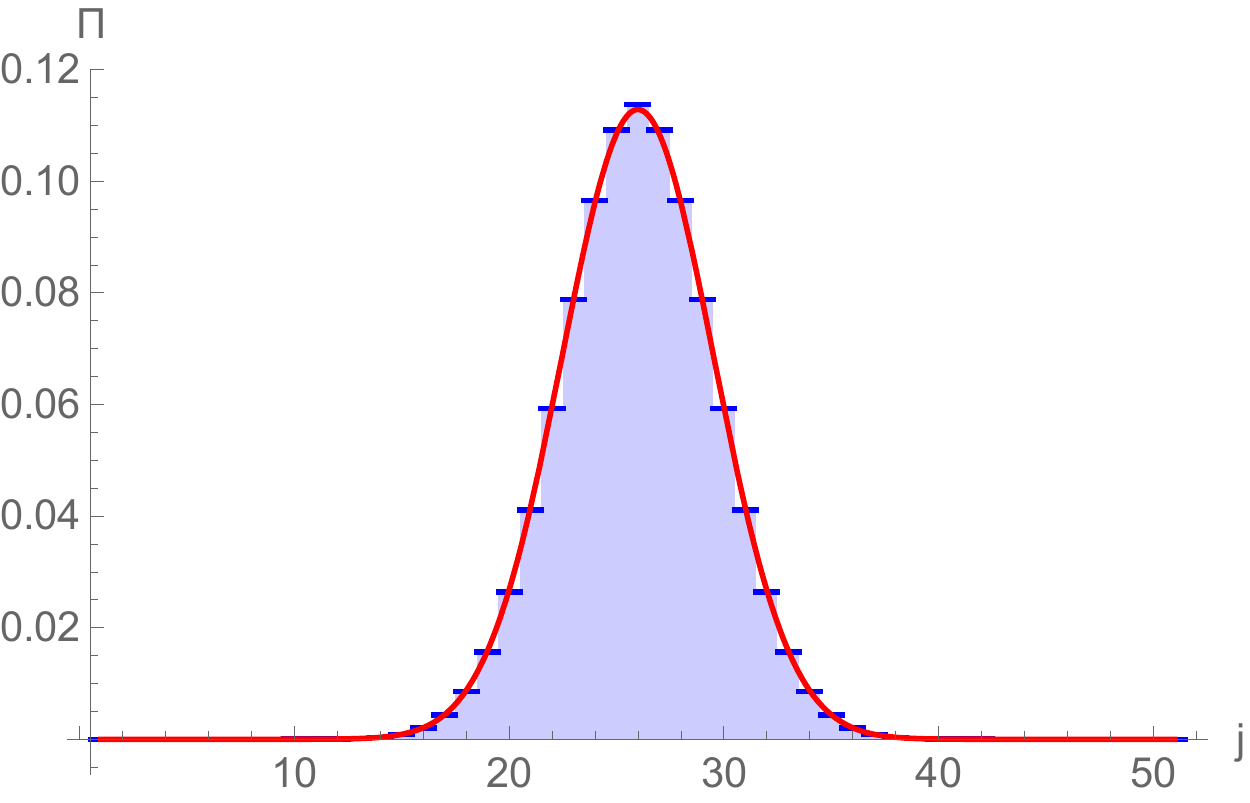}
\end{center}
\caption{
{\bf Exponentially localized interface states}. The blue curve presents the spatial extent of the localized states as defined by the metric (\ref{Loc}). The red curve represents a fit of the form (\ref{Loc-Approx}), displaying excellent agreement. 
}
\label{Fig-Loc}
\end{figure}

In this case, the symmetry breaking contribution to the full Hamiltonian may be written as
\be
H_\Delta = \sum_{\bk_\parallel, \, j} \Delta (j) \, \bs{c}_{\bk_\parallel, \, j}^\dagger \, \sigma_z \, \bs{c}_{\bk_\parallel, \, j},
\ee
where the dependence on $y$ is translated into the layer label $j$. As before, we are interested in a perturbation $\Delta(j)$ that changes sign at $y=0$. 
We consider the scenarios of both an even or odd number of layers. In the latter case we take $j_0 = \frac{1}{2} (M+1)$ so that the middle layer $j_0$ corresponds to $y=0$, implying that $\Delta$ vanishes at $y=0$. 
We consider a linear perturbation ranging from $-\Delta_0$ to $+ \Delta_0$, with $\Delta_0 >0$ that is given by
\be \label{Linear-Delta}
\Delta(j) = 2 \frac{j-1}{M-1} \Delta_0 - \Delta_0.
\ee
For an even number of layers, the gap function $\Delta$ given by (\ref{Linear-Delta}) does not vanish anywhere since there is no center layer. 
In Fig. \ref{Fig-EnergyBands} we display the corresponding energy dispersion for the different states with $\Delta_0 = 1$ and a total number of layers $M = 51$, as a function of $k_x$ (note that the solution is independent of $k_z$). The states plotted in red exhibit a gap-closing point at $k_x=n\pi$, implying that a line node is present. 
To establish the spatial extent of the nodal states, we introduce the following metric 
\be \label{Loc}
\Pi(j) = \frac{|\psi_j|^2}{{\bs \Psi}^\dagger\, \bs \Psi} 
\ee
where $\bs \Psi = (\psi_1, \, \psi_2 , \, \dots , \, \psi_j, \, \dots , \, \psi_M)^{\rm T}$ is the wave function of the state. 
The metric (\ref{Loc}) is shown in Fig. \ref{Fig-Loc}, revealing that the metallic interface state is exponentially localized to the center layer $j_0$. 

One may note that, after the introduction of $\Delta(j)$, the system is gapped in every layer but the one corresponding to $j_0$, where $\Delta(j_0) = 0$. It should therefore be expected that the interface states fall off as 
\be \label{Loc-Approx}
\frac{1}{c_0} e^{-\Delta(j) \, (j-j_0)},
\ee
where $c_0$ is a normalization constant. The red curve in Fig. \ref{Fig-Loc} shows a fit of the form (\ref{Loc-Approx}) with $c_0$ as a free parameter, revealing that there is an excellent agreement. The result for an even number of layers was found to be indistinguishable from the case of an odd number implying that the metallic interface states do not depend on details of the domain wall.

\section{Summary}
In conclusion, we have examined the stability of line-node semimetals in the presence of Coulomb interactions and found a chiral instability occurring at a finite interaction strength. The chiral order parameter exhibits a well-defined behavior in the limit of an infinite screening length despite the presence of infrared divergencies in this problem. 
By computing the Landau levels, we observe that an out-of-plane magnetic field couples to the chiral order parameter, implying that this degree of freedom can be controlled in experiments. 
While the drum-head edge states associated with line-node semimetals vanish in the chiral phase, we observe metallic interface-states in this regime, which exist on domain walls interpolating between regions of different chirality. These domain walls could conceivably be trapped on a sample with a concave geometry in experiments.

This work was supported by the Swedish Research Council (VR) through grant 2018-03882 and Stiftelsen Olle Engkvist via grant 204-0185. Computations were performed on resources provided by the Swedish National Infrastructure for Computing (SNIC) at the National Supercomputer Centre in Linköping, Sweden. J. C. would like to thank Lars Fritz for important input and discussions.

\bibliography{biblio.bib}

\end{document}